\documentclass[preprintnumbers,showpacs,amsmath,amssymb,floatfix,11pt,prd,onecolumn,
superscriptaddress,nofootinbib]{revtex4}
\usepackage{latexsym}
\usepackage{epsfig}
\usepackage{amssymb}

\begin{document}

\title{Wormholes in a Viable $f(T)$ Gravity}

\author{\bf Mubasher Jamil}
%\email{mjamil@camp.nust.edu.pk}
 \affiliation {Center for Advanced Mathematics and Physics (CAMP),
\\
National University of Sciences and Technology (NUST), H-12,
Islamabad, Pakistan} \affiliation {Eurasian International Center for
Theoretical Physics, Eurasian National University, Astana 010008,
Kazakhstan}

\author{\bf Davood Momeni}
%\email{d.momeni@yahoo.com}
 \affiliation{Eurasian International Center for Theoretical Physics,
Eurasian National University, Astana 010008, Kazakhstan}

\author{\bf Ratbay Myrzakulov}
%\email{rmyrzakulov@gmail.com}
 \affiliation{Eurasian International Center for Theoretical Physics,
 Eurasian National University, Astana 010008, Kazakhstan}

\begin{abstract}
{\bf Abstract:}  In this paper, we derive some new exact solutions
of static wormholes in $f(T)$ gravity. We discuss independent cases
of the pressure components including isotropic and anisotropic
pressure. Lastly we consider radial pressure satisfying a barotropic
equation of state. We also check the behavior of null energy
condition (NEC) for each case and observe that it is violated for
anisotropic while it is satisfied for isotropic and barotropic
cases.
\end{abstract}


\maketitle

\newpage
\section{Introduction}

Teleparallelism was proposed by Einstein as an attempt to unify
electromagnetism and gravity. But as time rolled on scientists lost
interest in this concept of unification and as a result, today
teleparallelism is just considered as an alternative classical gauge
theory for gravity. It corresponds to a diffeomorphism invariant
gauge model for the boost Poincare's  group \cite{trans}. The
crucial new idea, for Einstein, was the introduction of an
orthogonal basis, called the tetrads. Here the space-time is
characterized by a curvature-free linear connection, called the
Weitzenbock connection defining the scalar torsion without
Riemannian curvature. In  general relativity (GR), we use curvature
to unify the geometry and the gravitational field according to the
Mach's equivalence principle,  thus successfully describing the
gravitational interaction. Teleparallelism, explains gravitation by
torsion. Therefore it seems to be just a matter of convention that
whether gravity needs a curved spacetime. Exact solutions for this
newly proposed theory play a key role to test the validity of the
results as well as an attempt to find the correct path of the test
particles in the curved spacetime, based on the torsion. We are
focused in this paper on wormhole solutions, as a class of exact
solutions with a little different aspect. A typical spherically
symmetric spacetime can be described as a wormhole and as a bridge
between two distinct parts of the spacetime (see \cite{lobo23} for a
review).  It was first proposed by Morris and Thorne \cite{morris}
in the context of the time machine idea. In literature, wormholes
have been discussed
 \cite{jamil},\cite{kuhf},\cite{luis}. Otherwise, to explain the new challenge of the cosmology,
the modified gravity has gained enormous popularity in the
cosmological society since it passes several solar system and
astrophysical tests successfully \cite{sergei2}. In addition to
this, modified gravity models can be used as well, for the recent
cosmic acceleration without resorting to any form of DE. For example
some popular models like Loop quantum gravity \cite{lqc}, extra
dimensional braneworlds \cite{brane}, $f(R)$ \cite{fr}, $f(T)$
\cite{ft, miao, miaoli, sotirio}. \\
In this paper,  we derive some new exact solutions of static
wormholes in $f(T)$ gravity. We discuss independent cases of the
pressure components including isotropic and anisotropic pressure.
Lastly we consider the non zero radial pressure $p_r(r)$ satisfying
a typical barotropic equation of state. We also check the behavior
of null energy condition (NEC) for each case and observe that it is
violated for anisotropic while it is satisfied for isotropic and
barotropic cases.

The plan of this paper is as follows.  In section II, the basic
equations for the $f(T)$ model is discussed . In section III, the
field equations are written. In section IV, wormhole solutions in
$f(T)$ gravity are discussed. In section V, we propose our viable
$f(T)$ model. In section VI, an exact anisotropic solution is
derived. In section VII, the isotropic solutions are derived. In
section VIII, the barotropic solutions are investigated. Finally we
end with some concluding remarks in section IX.

\section{Gravitational field equations in a f(T) gravity}

We assume that the metric is static spherically symmetric in the
usual Schwarzschild coordinates in the following form \cite{lobo},
\begin{equation}
ds^2 = e^{a(r)} dt^2 - e^{b(r)} dr^2 -r^2 (d\theta^2 + \sin^2\theta
d\varphi^2),
  \label{metric1off}
\end{equation}
Here $a(r)$ and $b(r)$ are the unknown components of the  diagonal
metric. We take the orthogonal but off-diagonal tetrads to avoid the
teleparallel gravity (TEGR) with $f_{TT}=0$, given by \cite{lobo}
\begin{eqnarray}
  e^i{}_{\mu} = \left(
\begin{array}{cccc}
  e^{a/2} & 0 & 0 & 0 \\
  0 & e^{b/2} \sin\theta\cos\phi  & r\cos \theta\cos\phi & -r\sin \theta\sin \phi \\
  0 & e^{b/2} \sin\theta\sin\phi & r\cos \theta\sin\phi& r\sin\theta \cos\phi \\
  0 & e^{b/2} \cos \theta  & -r \sin\theta & 0
\end{array}
\right).
  \label{tetradoff}\nonumber
\end{eqnarray}
%The off diagonal basis tetrad is related to its diagonal form by doing a rotation.

Also $det(e^i{}_\mu)$ reads $e=exp((a+b)/2)r^2\sin\theta$. Further the  scalar torsion $T$ is
\begin{eqnarray}
  T(r) &=& \frac{2 e^{-b} (e^{b/2}-1)(e^{b/2}-1-r a ')}{r^2},
  \label{Tscalaroff}
\end{eqnarray}
respectively.\\
 The equations of motion are the following \cite{lobo}
\begin{eqnarray}
  4\pi\rho(r) &=& \frac{e^{-b/2}}{r}(1-e^{-b/2}) T'f_{TT}-\Big(\frac{T}{4}-\frac{1}
  {2r^2}\Big)f_T+\frac{e^{-b}}{2 r^2} \Big(rb '-1\Big)f_T-\frac{f}{4},
  \label{field:toff}\\
  4\pi p_r(r) &=&\Big[- \frac{1}{2r^2}+\frac{T}{4}+\frac{e^{-b}}{2r^2}
  (1+ra')\Big]f_T-\frac{f}{4},
  \label{field:roff}\\
  4\pi p_t(r) &=& \frac{e^{-b}}{2}\Big(\frac{a'}{2}+\frac{1}{r}-\frac{e^{b/2}}{r}\Big)T'
  f_{TT}+ f_T\Big\{ \frac{T}{4}+\frac{e^{-b}}{2 r} \Big[\Big(\frac{1}
 {2}+\frac{ra'}{4}\Big) \Big(a'-b'\Big)+\frac{ra''}{2}\Big]\Big\}-
 \frac{f}{4},
  \label{field:thetaoff}
\end{eqnarray}
where $\rho(r), p_r(r),p_t(r)$ are the energy density, the radial
pressure and the pressure. The above system of field
equations (\ref{field:toff})-(\ref{field:thetaoff}) is not closed. Consequently we fix some of these functions.

\section{Wormhole solutions in $f(T)$ gravity}

For a better understanding of numerous features of the wormhole
geometry, we shall use the following  redefinition of the metric
function
\begin{equation}
e ^{-b(r)} = 1-\frac{\beta(r)}{r} . \label{metricwormhole}
\end{equation}
In the theory of wormholes, $a(r)$ and $\beta(r)$ are arbitrary
functions of the radial coordinate. In brief in the $3+1$
decomposition language, $a(r)$ denotes the redshift function, which
is related to the gravitational redshift $g_{00}$. Also $\beta(r)$
defines the shape function and it specifies the geometrical shape of
the stationary wormhole \cite{Morris}. The coordinate $r$  is in the
strip $r_0<r<+\infty,$ $r_0$ denotes the radius of wormhole's
throat. By flaring out of the throat, we mean to impose the
condition $(\beta-\beta' r)/2\beta^2>0$~\cite{Morris} and at the
throat location we observe that $\beta '(r_0)<1$.

%The flaring out condition of the throat is a fundamental property in
%wormholes, and through the Einstein field equations it was found
%that some of these solutions possess a peculiar property, namely
%exotic matter, involving a stress-energy tensor that violates the
%null energy condition. One good example for such fluids is dark
%energy with EoS $p\approx-\rho$.
It turns out that that Morris-Thorne wormholes disobey all energy
conditions and averaged null energy condition. In fact, the weak
energy condition (WEC) proposes that the energy density must remain
locally positive definite, i.e.
$$
\rho=T_{\mu\nu}U^\mu U^\nu \geq 0,
$$
for all timelike vector fields $U^\mu$, where $T_{\mu\nu}$ is the
usual energy-momentum (EM) tensor. In the limit of the quantum
effects, this expression of the EM tensor can be replaced by the
semi classical quantum expectation value  $<T^{vaccua}_{\mu\nu}>$,
where it can be replaced as a source term in the right hand side of
the gravitational field equation. For example, this  kind of EM
tensors with conformal anomaly can produce a rich family of the
exact (A)dS black holes in the context of the $f(R)$ gravity
\cite{hendi}.

Further we can show that by replacing the general expression of the
EM tensor for a typical Lorentz invariance Lagrangian $\mathcal{L}$,
by the expression
$$
T_{\mu\nu}=g_{\mu\nu}\mathcal{L}-2\frac{\delta \mathcal{L}}{\delta
g^{\mu\nu}},
$$
and by adopting a timelike velocity vector stream by
$U^{\mu}=\phi^{;\mu}$, again this new expression remains positive
definite. In the locally Lorentzian frame of the matter, for example
in the co moving frame in the perfect fluids, this means $\rho>0$
and $\rho+p_i\geq0$, where $i=r,t$. The WEC implies NEC, i.e. the
null energy condition: $T_{\mu\nu}k^\mu k^\nu \geq 0$, here $k^\mu$
denotes a null vector field \cite{ellis}. Our goal is to find some
exact solutions for wormholes and check whether they satisfy or
violate the above energy conditions. We briefly  review the usual
energy conditions in the next section.

\section{Energy conditions }

The energy conditions are used in different contexts to derive
variety of general results which hold for different physical situations.
Under these conditions, one allows not just gravity to be attractive
but also the energy density to be positive and flows not to be
faster than light \cite{wangg}. The notion of energy conditions
arise from the Raychaudhuri equation, given by
\begin{equation}
\frac{d\theta}{d\tau}=-\frac{1}{3}\theta^2-\sigma_{\mu\nu}\sigma^{\mu\nu}
+\omega_{\mu\nu}\omega^{\mu\nu}-R_{\mu\nu}u^\mu u^\nu,
\end{equation}
where $u^\mu$ is a vector field representing the congruence of
timelike geodesics. Also $R_{\mu\nu}$, represent Ricci tensor
$\theta$,the expansion parameter, $\sigma_{\mu\nu}$ the shear and
$\omega_{\mu\nu}$, the rotation associated with the congruence
respectively. For a congruence of null geodesics, $k^\mu$, we get
\begin{equation}
\frac{d\theta}{d\tau}=-\frac{1}{2}\theta^2-\sigma_{\mu\nu}\sigma^{\mu\nu}
+\omega_{\mu\nu}\omega^{\mu\nu}-R_{\mu\nu}k^\mu k^\nu,
\end{equation}
%From both the Raychaudhuri equations, it is apparent that these are
%purely geometric and independent of any gravity theory.
which is called the Raychaudhuri equation. The essence of such
checking energy conditions is independent of any gravity theory and
that these are purely geometrical (for a review on the energy
conditions, see the classic text \cite{hawking}).\\
The null energy condition (NEC), weak energy condition (WEC), strong
energy condition (SEC) and the dominant energy condition (DEC) are
respectively given by \cite{lobo,anzhong}
\begin{eqnarray}
\text{NEC}&\Longleftrightarrow&\rho_{\text{eff}}+p_{\text{eff}}\geq0.\label{n1}\\
\text{WEC}&\Longleftrightarrow& \rho_{\text{eff}}\geq0\ \text{and}\ \rho_{\text{eff}}+p_{\text{eff}}\geq0.\label{n2}\\
\text{SEC}&\Longleftrightarrow& \rho_{\text{eff}}+3p_{\text{eff}}\geq0\ \text{and}\ \rho_{\text{eff}}+p_{\text{eff}}\geq0.\label{n3}\\
\text{DEC}&\Longleftrightarrow& \rho_{\text{eff}}\geq0\ \text{and}\
\rho_{\text{eff}}\pm p_{\text{eff}}\geq0,\label{n4}
\end{eqnarray}
where $\rho_{\text{eff}}$ and $p_{\text{eff}}$ are respectively the
effective energy density and pressure. The matter supporting the
wormhole geometry is termed ``exotic'' matter if it violates the
above energy conditions, particularly the NEC, as its violation
leads to the violation of all the remaining energy conditions
\cite{krasnikov}.

\section{Equations of motion for a viable model of $f(T)$}

We must fix the form of the $f(T)$ to simplify the equations of
motion and to find a class of the solutions, corresponds to the
wormhole. Recently we proposed a physically reasonable expression
for $f(T)$  \cite{attractor} in the following form,
\begin{equation}
f(T)=2c_1 \sqrt{-T} +\alpha T+c_2\label{ft},
\end{equation}
Here $\{\alpha,c_1,c_2\}$ are arbitrary constants but the values
must be obtain by some additional cosmological evidences.

Now  field equations~(\ref{field:toff})--(\ref{field:thetaoff}) and
(\ref{ft}) give us the following system of equation of motion

\begin{eqnarray}
  4\pi \rho(r)&=&\frac{1}{8}{{\rm e}^{-b}} \Big( -1+{{\rm e}^{-\frac{1}{2}b}} \Big) \sqrt {2}{
{\rm e}^{-\frac{1}{2}b}} \Big( -4  {{\rm e}^{b}}  +
 \Big( -2{r}^{2}{\it a''}+ \Big( 2r+{\it b'}{r}^{2} \Big) {
\it a'}+8+2r{\it b'} \Big) {{\rm e}^{\frac{1}{2}b}}+2{r}^{2}{\it a''}\nonumber\\&& +
 \Big( -2{\it b'}{r}^{2}-2r \Big) {\it a'}-2r{\it b'}-4
 \Big) c_{{1}} \Big( {\frac {{{\rm e}^{-b}} \Big( {{\rm e}^{\frac{1}{2}b
}}-1 \Big)  \Big( {{\rm e}^{\frac{1}{2}b}}-1-r{\it a'} \Big) }{{r}^{2}} }
\Big) ^{-\frac{3}{2}}{r}^{-3} \nonumber\\&& - \Big( \frac{1}{2}c_{{1}}\sqrt
{2}{\frac {1}{\sqrt {{\frac {{{\rm e}^{-b}}
 \Big( {{\rm e}^{\frac{1}{2}b}}-1 \Big)  \Big( {{\rm e}^{\frac{1}{2}b}}-1-r{
\it a'} \Big) }{{r}^{2}}}}}}+\alpha \Big)  \Big( \frac{1}{2}{\frac {{ {\rm
e}^{-b}} \Big( {{\rm e}^{\frac{1}{2}b}}-1 \Big)  \Big( {{\rm e}^{\frac{1}{2}b}}-1-r{\it a'} \Big) }{{r}^{2}}}-\frac{1}{2}{r}^{-2} \Big)
 \nonumber\\&& -\frac{1}{2}c_{{1}}\sqrt {2}\sqrt {{\frac {{{\rm e}^{-b}} \Big( {{\rm e}^{\frac{1}{2}b}}-1 \Big)  \Big( {{\rm e}^{\frac{1}{2}b}}-1-r{\it a'} \Big) }{{r}
^{2}}}}-\frac{1}{2}{\frac {\alpha{{\rm e}^{-b}} \Big( {{\rm e}^{\frac{1}{2}b}}- 1
\Big)  \Big( {{\rm e}^{\frac{1}{2}b}}-1-r{\it a'} \Big) }{{r}^{2}}}
\\&&\nonumber-\frac{1}{4}c_{{2}}+\frac{1}{4}{{\rm e}^{-b}} \Big( c_{{1}}\sqrt {2}{\frac {1}{
\sqrt {{\frac {{{\rm e}^{-b}} \Big( {{\rm e}^{\frac{1}{2}b}}-1 \Big)
 \Big( {{\rm e}^{\frac{1}{2}b}}-1-r{\it a'} \Big) }{{r}^{2}}}}}}+2
\alpha \Big)  \Big( r{\it b'}-1 \Big) {r}^{-2},
\end{eqnarray}

\begin{eqnarray}
 4\pi p_r(r)&=&-\frac{1}{4} \Big(  \Big(  \Big( -2\alpha-2\alpha r{\it a'} \Big)
{{\rm e}^{-b}}+{r}^{2}c_{{2}}+2\alpha \Big) \sqrt {{\frac {{ {\rm
e}^{-b}} \Big( {{\rm e}^{\frac{1}{2}b}}-1 \Big)  \Big( {{\rm e}^{\frac{1}{2}b}}-1-r{\it a'} \Big) }{{r}^{2}}}} \nonumber\\&&  - \Big( -1- \Big(
{{\rm e}^{\frac{1}{2}b}} \Big) ^{2}{{\rm e}^{-b}}+{{\rm e}^{-b}} \Big(
2+r{\it a'}
 \Big) {{\rm e}^{\frac{1}{2}b}} \Big) \sqrt {2}c_{{1}} \Big) {\frac {1
}{\sqrt {{\frac {{{\rm e}^{-b}} \Big( {{\rm e}^{\frac{1}{2}b}}-1 \Big)
 \Big( {{\rm e}^{\frac{1}{2}b}}-1-r{\it a'} \Big)
 }{{r}^{2}}}}}}{r}^{-2},\nonumber\\
\end{eqnarray}

\begin{eqnarray}
4\pi p_t(r)&=&\frac{1}{8}{\frac {1}{\sqrt {{\frac {{{\rm e}^{-b}} \Big(
{{\rm e}^{\frac{1}{2}b} }-1 \Big)  \Big( {{\rm e}^{\frac{1}{2}b}}-1-r{\it a'} \Big)
}{{r}^{2}}} }}}{r}^{-2}\times\Big(r \Big( \alpha \Big( 2r{\it a''}+
\Big( {\it a'}-{\it b'}
 \Big)  \Big( 2+r{\it a'} \Big)  \Big) {{\rm e}^{-b}}-2rc_{{2
}} \Big) \nonumber\\&&\times\sqrt {{\frac {{{\rm e}^{-b}} \Big(
{{\rm e}^{1/2b}}-1
 \Big)  \Big( {{\rm e}^{\frac{1}{2}b}}-1-r{\it a'} \Big) }{{r}^{2}}}}+\frac{1}{2} \Big( -4  {{\rm e}^{b}} + \Big( 4r{ \it a'}+8
\Big) {{\rm e}^{\frac{1}{2}b}}+2{r}^{2}{\it a''}\nonumber\\&&+{r}^{2}{{\it
a'}}^{2}+ \Big( -2r-{\it b'}{r}^{2} \Big) {\it a'}-2r{\it b'} -4
\Big) {{\rm e}^{-b}}\sqrt {2}c_{{1}} \Big).
\end{eqnarray}

Now we will discuss the possible physical solution in the following
three cases.
\begin{enumerate}
\item \textbf{Anisotropic fluid}:  $p_r(r)\neq p_t(r)$.
\item \textbf{Isotropic fluid}:  $p_r(r)=p_t(r)$.
\item \textbf{Barotropic EoS}:  $p_r(r)=k\rho(r)$.
\end{enumerate}

\section{anisotropic solution }
In the first class for exact solution with anisotropic components in
the  EM tensor of the matter fields,We assume that \cite{lobo}
\begin{equation}\label{f}
  a(r)=c_1, \qquad b(r)=-\log  \Big( 1- \Big( {\frac {r_{{0}}}{r}} \Big) ^{n+1}
  \Big),
\end{equation}
where $c_1$, $n$ and $r_0$ are positive constants. Inserting these
functions (\ref{f}) into the EM tensor system,
Eqs.~(\ref{field:toff})--(\ref{field:thetaoff}), provides the
following solutions
\begin{eqnarray}
  4\pi \rho( r )& =&\frac{1}{4}\Big( c_{{1}}\sqrt {2}
 \Big( {\frac {r_{{0}}}{r}} \Big) ^{n+1}nr+3c_{{1}}\sqrt {2}
 \Big( {\frac {r_{{0}}}{r}} \Big) ^{n+1}-2\alpha\sqrt { \Big(
-1+\sqrt {1- \Big( {\frac {r_{{0}}}{r}} \Big) ^{n+1}} \Big) ^{2}{
r}^{-2}} \Big( {\frac {r_{{0}}}{r}} \Big)
^{n+1}n\nonumber\\&&+4\alpha\sqrt { \Big( -1+\sqrt {1- \Big( {\frac
{r_{{0}}}{r}} \Big) ^{n+1}}
 \Big) ^{2}{r}^{-2}} \Big( {\frac {r_{{0}}}{r}} \Big) ^{n+1}+3c
_{{1}}\sqrt {2}r \Big( {\frac {r_{{0}}}{r}} \Big) ^{n+1}-c_{{1}}
\sqrt {2} \Big( {\frac {r_{{0}}}{r}} \Big)
^{n+1}n\nonumber\\&&-8\alpha \sqrt { \Big( -1+\sqrt {1- \Big( {\frac
{r_{{0}}}{r}} \Big) ^{n+1} } \Big) ^{2}{r}^{-2}}-6c_{{1}}\sqrt
{2}+8\alpha\sqrt { \Big( -1+\sqrt {1- \Big( {\frac {r_{{0}}}{r}}
\Big) ^{n+1}} \Big) ^{2}{ r}^{-2}}\nonumber\\&&\times\sqrt {1- \Big(
{\frac {r_{{0}}}{r}} \Big) ^{n+1}}-2c_{{1 }}\sqrt {2}r+6\sqrt {1-
\Big( {\frac {r_{{0}}}{r}} \Big) ^{n+1}}c _{{1}}\sqrt
{2}-c_{{2}}{r}^{2}\sqrt { \Big( -1+\sqrt {1- \Big( { \frac
{r_{{0}}}{r}} \Big) ^{n+1}} \Big) ^{2}{r}^{-2}}\nonumber\\&&+2\sqrt
{1-
 \Big( {\frac {r_{{0}}}{r}} \Big) ^{n+1}}c_{{1}}\sqrt {2}r \Big)
{r}^{-2}{\frac {1}{\sqrt { \Big( -1+\sqrt {1- \Big( {\frac {r_{{0}}}
{r}} \Big) ^{n+1}} \Big) ^{2}{r}^{-2}}}},
\end{eqnarray}
\begin{eqnarray}
4\pi p_{{r}} ( r)& =&-\frac{1}{4} \Big( 2\alpha\sqrt { \Big( -1+\sqrt {1-
\Big( {\frac {r_{{0}}}{r}} \Big) ^{n+1}} \Big) ^{2}{r}^{-2}} \Big( {
\frac {r_{{0}}}{r}} \Big) ^{n+1}\sqrt {1- \Big( {\frac {r_{{0}}}{r}
} \Big) ^{n+1}}\nonumber\\&&+\sqrt { \Big( -1+\sqrt {1- \Big( {\frac
{r_{{0}}}{ r}} \Big) ^{n+1}} \Big) ^{2}{r}^{-2}}\sqrt {1- \Big(
{\frac {r_{{0 }}}{r}} \Big) ^{n+1}}c_{{2}}{r}^{2}+2\sqrt {1- \Big(
{\frac {r_{{0 }}}{r}} \Big) ^{n+1}}c_{{1}}\sqrt {2}-2c_{{1}}\sqrt
{2}\nonumber\\&&+2c_{{1}} \sqrt {2} \Big( {\frac {r_{{0}}}{r}} \Big)
^{n+1} \Big) {\frac {1 }{\sqrt { \Big( -1+\sqrt {1- \Big( {\frac
{r_{{0}}}{r}} \Big) ^{n+ 1}} \Big) ^{2}{r}^{-2}}}}{\frac {1}{\sqrt
{1- \Big( {\frac {r_{{0}} }{r}} \Big) ^{n+1}}}}{r}^{-2},
\end{eqnarray}
\begin{eqnarray}
 4\pi p_{{t}} ( r ) &=&\frac{1}{8}
\Big( 2\alpha\sqrt { \Big( -1+\sqrt {1- \Big( {\frac {r_{{0} }}{r}}
\Big) ^{n+1}} \Big) ^{2}{r}^{-2}} \Big( {\frac {r_{{0}}}{r }} \Big)
^{n+1}\sqrt {1- \Big( {\frac {r_{{0}}}{r}} \Big) ^{n+1}}
n\nonumber\\&&+2\alpha\sqrt { \Big( -1+\sqrt {1- \Big( {\frac
{r_{{0}}}{r}}
 \Big) ^{n+1}} \Big) ^{2}{r}^{-2}} \Big( {\frac {r_{{0}}}{r}}
 \Big) ^{n+1}\sqrt {1- \Big( {\frac {r_{{0}}}{r}} \Big) ^{n+1}}\nonumber\\&&-2
\sqrt { \Big( -1+\sqrt {1- \Big( {\frac {r_{{0}}}{r}} \Big) ^{n+ 1}}
\Big) ^{2}{r}^{-2}}\sqrt {1- \Big( {\frac {r_{{0}}}{r}}
 \Big) ^{n+1}}c_{{2}}{r}^{2}-4\sqrt {1- \Big( {\frac {r_{{0}}}{r}
} \Big) ^{n+1}}c_{{1}}\sqrt {2}+4c_{{1}}\sqrt
{2}\nonumber\\&&-4c_{{1}}\sqrt {2} \Big( {\frac {r_{{0}}}{r}} \Big)
^{n+1}+c_{{1}}\sqrt {2}
 \Big( {\frac {r_{{0}}}{r}} \Big) ^{n+1}n\sqrt {1- \Big( {\frac {r
_{{0}}}{r}} \Big) ^{n+1}}+3\sqrt {1- \Big( {\frac {r_{{0}}}{r}}
 \Big) ^{n+1}}c_{{1}}\sqrt {2} \Big( {\frac {r_{{0}}}{r}} \Big) ^
{n+1} \Big)\nonumber\\&&\times {\frac {1}{\sqrt { \Big( -1+\sqrt {1-
\Big( {\frac {r_ {{0}}}{r}} \Big) ^{n+1}} \Big)
^{2}{r}^{-2}}}}{\frac {1}{\sqrt {1-
 \Big( {\frac {r_{{0}}}{r}} \Big) ^{n+1}}}}{r}^{-2}.
\end{eqnarray}
The energy conditions read
\begin{eqnarray}
\rho+p_r&=&\frac{1}{16\pi}{\frac {1}{\sqrt { \Big( -1+\sqrt {1-
\Big( {\frac {r_{{0}}}{r}} \Big) ^{n+1}} \Big) ^{2}{r}^{-2}}}}{\frac
{1}{\sqrt {1- \Big( {\frac {r_{{0}}}{r}} \Big) ^{n+1}}}}{r}^{-2}
\Big(\sqrt {1- \Big( {\frac {r_{{0}}}{r}} \Big) ^{n+1}}c_{{1}}\sqrt
{2}
 \Big( {\frac {r_{{0}}}{r}} \Big) ^{n+1}nr\nonumber\\&&+3\sqrt {1- \Big( {
\frac {r_{{0}}}{r}} \Big) ^{n+1}}c_{{1}}\sqrt {2} \Big( {\frac {r_{
{0}}}{r}} \Big) ^{n+1}-2\sqrt {1- \Big( {\frac {r_{{0}}}{r}}
 \Big) ^{n+1}}\alpha\sqrt { \Big( -1+\sqrt {1- \Big( {\frac {r_{
{0}}}{x}} \Big) ^{n+1}} \Big) ^{2}{r}^{-2}} \Big( {\frac {r_{{0}}
}{r}} \Big) ^{n+1}n \nonumber\\&& +2\sqrt {1- \Big( {\frac
{r_{{0}}}{r}}
 \Big) ^{n+1}}\alpha\sqrt { \Big( -1+\sqrt {1- \Big( {\frac {r_{
{0}}}{r}} \Big) ^{n+1}} \Big) ^{2}{r}^{-2}} \Big( {\frac {r_{{0}}
}{r}} \Big) ^{n+1}+3\sqrt {1- \Big( {\frac {r_{{0}}}{r}} \Big)
^{n+1}}c_{{1}}\sqrt {2}r \Big( {\frac {r_{{0}}}{r}} \Big)
^{n+1}\nonumber\\&&- \sqrt {1- \Big( {\frac {r_{{0}}}{r}} \Big)
^{n+1}}c_{{1}}\sqrt {2}
 \Big( {\frac {r_{{0}}}{r}} \Big) ^{n+1}n-8\sqrt {1- \Big( {
\frac {r_{{0}}}{r}} \Big) ^{n+1}}\alpha\sqrt { \Big( -1+\sqrt {1-
 \Big( {\frac {r_{{0}}}{r}} \Big) ^{n+1}} \Big) ^{2}{r}^{-2}} \nonumber\\&& -8
\sqrt {1- \Big( {\frac {r_{{0}}}{r}} \Big) ^{n+1}}c_{{1}}\sqrt {2}+
8\alpha\sqrt { \Big( -1+\sqrt {1- \Big( {\frac {r_{{0}}}{r}}
 \Big) ^{n+1}} \Big) ^{2}{r}^{-2}}\nonumber\\&&-8\alpha\sqrt { \Big( -1+
\sqrt {1- \Big( {\frac {r_{{0}}}{r}} \Big) ^{n+1}} \Big) ^{2}{r}^
{-2}} \Big( {\frac {r_{{0}}}{r}} \Big) ^{n+1}-2\sqrt {1- \Big( {
\frac {r_{{0}}}{r}} \Big) ^{n+1}}c_{{1}}\sqrt {2}r+8c_{{1}}\sqrt {
2}\nonumber\\&&-8c_{{1}}\sqrt {2} \Big( {\frac {r_{{0}}}{r}} \Big)
^{n+1}-2 \sqrt { \Big( -1+\sqrt {1- \Big( {\frac {r_{{0}}}{r}} \Big)
^{n+1} } \Big) ^{2}{r}^{-2}}\sqrt {1- \Big( {\frac {r_{{0}}}{r}}
\Big) ^ {n+1}}c_{{2}}{r}^{2}\nonumber\\&&+2c_{{1}}\sqrt
{2}r-2c_{{1}}\sqrt {2}r \Big( {\frac {r_{{0}}}{r}} \Big) ^{n+1}
\Big).\label{nec1}
\end{eqnarray}

\begin{eqnarray}
\rho+p_t&=&\frac{1}{32\pi}{\frac {1}{\sqrt { \Big( -1+\sqrt {1-
\Big( {\frac {r_{{0}}}{r}} \Big) ^{n+1}} \Big) ^{2}{r}^{-2}}}}{\frac
{1}{\sqrt {1- \Big( {\frac {r_{{0}}}{r}} \Big)
^{n+1}}}}{r}^{-2}\Big(2\sqrt {1- \Big( {\frac {r_{{0}}}{r}} \Big)
^{n+1}}c_{{1}}\sqrt { 2} \Big( {\frac {r_{{0}}}{r}} \Big)
^{n+1}nr\nonumber\\&&+6\sqrt {1- \Big( { \frac {r_{{0}}}{r}} \Big)
^{n+1}}c_{{1}}\sqrt {2} \Big( {\frac {r_{ {0}}}{r}} \Big)
^{n+1}-4\sqrt {1- \Big( {\frac {r_{{0}}}{r}}
 \Big) ^{n+1}}\alpha\sqrt { \Big( -1+\sqrt {1- \Big( {\frac {r_{
{0}}}{r}} \Big) ^{n+1}} \Big) ^{2}{r}^{-2}} \Big( {\frac {r_{{0}}
}{r}} \Big) ^{n+1}n\nonumber\\&&+8\sqrt {1- \Big( {\frac
{r_{{0}}}{r}}
 \Big) ^{n+1}}\alpha\sqrt { \Big( -1+\sqrt {1- \Big( {\frac {r_{
{0}}}{r}} \Big) ^{n+1}} \Big) ^{2}{r}^{-2}} \Big( {\frac {r_{{0}}
}{r}} \Big) ^{n+1}+6\sqrt {1- \Big( {\frac {r_{{0}}}{r}} \Big)
^{n+1}}c_{{1}}\sqrt {2}r \Big( {\frac {r_{{0}}}{r}} \Big)
^{n+1}\nonumber\\&&-2 \sqrt {1- \Big( {\frac {r_{{0}}}{r}} \Big)
^{n+1}}c_{{1}}\sqrt {2 } \Big( {\frac {r_{{0}}}{r}} \Big)
^{n+1}n-16\sqrt {1- \Big( { \frac {r_{{0}}}{r}} \Big)
^{n+1}}\alpha\sqrt { \Big( -1+\sqrt {1-
 \Big( {\frac {r_{{0}}}{r}} \Big) ^{n+1}} \Big) ^{2}{r}^{-2}}\nonumber\\&&-12
\sqrt {1- \Big( {\frac {r_{{0}}}{r}} \Big) ^{n+1}}c_{{1}}\sqrt {2
}+16\alpha\sqrt { \Big( -1+\sqrt {1- \Big( {\frac {r_{{0}}}{r}}
 \Big) ^{n+1}} \Big) ^{2}{r}^{-2}}\nonumber\\&&-16\alpha\sqrt { \Big( -1+
\sqrt {1- \Big( {\frac {r_{{0}}}{r}} \Big) ^{n+1}} \Big) ^{2}{r}^
{-2}} \Big( {\frac {r_{{0}}}{r}} \Big) ^{n+1}-4\sqrt {1- \Big( {
\frac {r_{{0}}}{r}} \Big) ^{n+1}}c_{{1}}\sqrt {2}r+12c_{{1}}\sqrt
{2}\nonumber\\&&-12c_{{1}}\sqrt {2} \Big( {\frac {r_{{0}}}{r}} \Big)
^{n+1}-2 \sqrt { \Big( -1+\sqrt {1- \Big( {\frac {r_{{0}}}{r}} \Big)
^{n+ 1}} \Big) ^{2}{r}^{-2}}\sqrt {1- \Big( {\frac {r_{{0}}}{r}}
 \Big) ^{n+1}}c_{{2}}{r}^{2}+4c_{{1}}\sqrt {2}r\nonumber\\&&-4c_{{1}}\sqrt {2
}r \Big( {\frac {r_{{0}}}{r}} \Big) ^{n+1}+2\pi \sqrt {1-
 \Big( {\frac {r_{{0}}}{r}} \Big) ^{n+1}}\alpha\sqrt { \Big( -1+
\sqrt {1- \Big( {\frac {r_{{0}}}{r}} \Big) ^{n+1}} \Big) ^{2}{r}^
{-2}} \Big( {\frac {r_{{0}}}{r}} \Big) ^{n+1}n\nonumber\\&&+2\pi
\sqrt {1-
 \Big( {\frac {r_{{0}}}{r}} \Big) ^{n+1}}\alpha\sqrt { \Big( -1+
\sqrt {1- \Big( {\frac {r_{{0}}}{r}} \Big) ^{n+1}} \Big) ^{2}{r}^
{-2}} \Big( {\frac {r_{{0}}}{r}} \Big) ^{n+1}\nonumber\\&&-2\pi
\sqrt {
 \Big( -1+\sqrt {1- \Big( {\frac {r_{{0}}}{r}} \Big) ^{n+1}}
 \Big) ^{2}{r}^{-2}}\sqrt {1- \Big( {\frac {r_{{0}}}{r}} \Big) ^{
n+1}}c_{{2}}r^{2}-4\pi \sqrt {1- \Big( {\frac {r_{{0}}}{r}}
 \Big) ^{n+1}}c_{{1}}\sqrt {2}\nonumber\\&&+4\pi c_{{1}}\sqrt {2}-4\pi c_
{{1}}\sqrt {2} \Big( {\frac {r_{{0}}}{r}} \Big) ^{n+1}+\pi \sqrt {1-
\Big( {\frac {r_{{0}}}{r}} \Big) ^{n+1}}c_{{1}}\sqrt {2}
 \Big( {\frac {r_{{0}}}{r}} \Big) ^{n+1}n\nonumber\\&&+3\pi \sqrt {1-
 \Big( {\frac {r_{{0}}}{r}} \Big) ^{n+1}}c_{{1}}\sqrt {2} \Big( {
\frac {r_{{0}}}{r}} \Big) ^{n+1} \Big)\label{nec2}.
\end{eqnarray}

The qualitative behavior of the energy
density;  the radial pressure; and the
 tangential pressure is plotted in Fig.1, left panel. Note that the
energy density is negative in entire spacetime. So the anisotropic
case violates the WEC also NEC.  But just for a small region which
is denoted by $r<0.25$, the energy density is positive. So, just
there is one possibility to have a \textit{micro} or \textit{tiny}
wormholes but in any case as we observe in the right panel, the  NEC
violated. The NEC for  $(p_t,p_r)$ must be check by positivity  of
the (\ref{nec1},\ref{nec2}). The violation of the NEC is for both
cases of the radial and the tangential components of the pressure.
Note that we adopted the parameters of our $f(T)$ to meet the
observational data by \textit{cosmography} technique.

\begin{figure*}[thbp]
\begin{tabular}{rl}
\includegraphics[width=7.5cm]{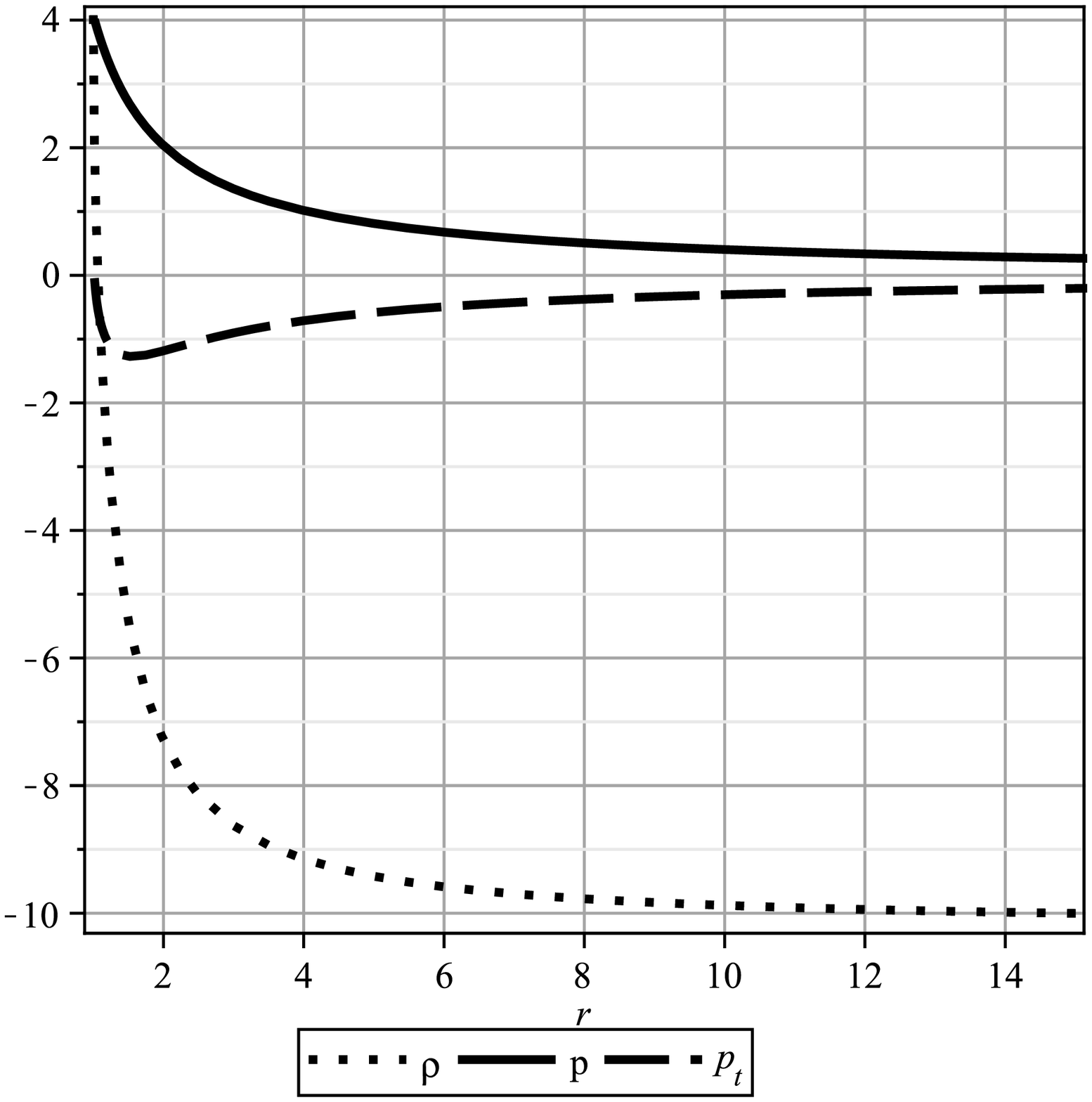}&
\includegraphics[width=7.5cm]{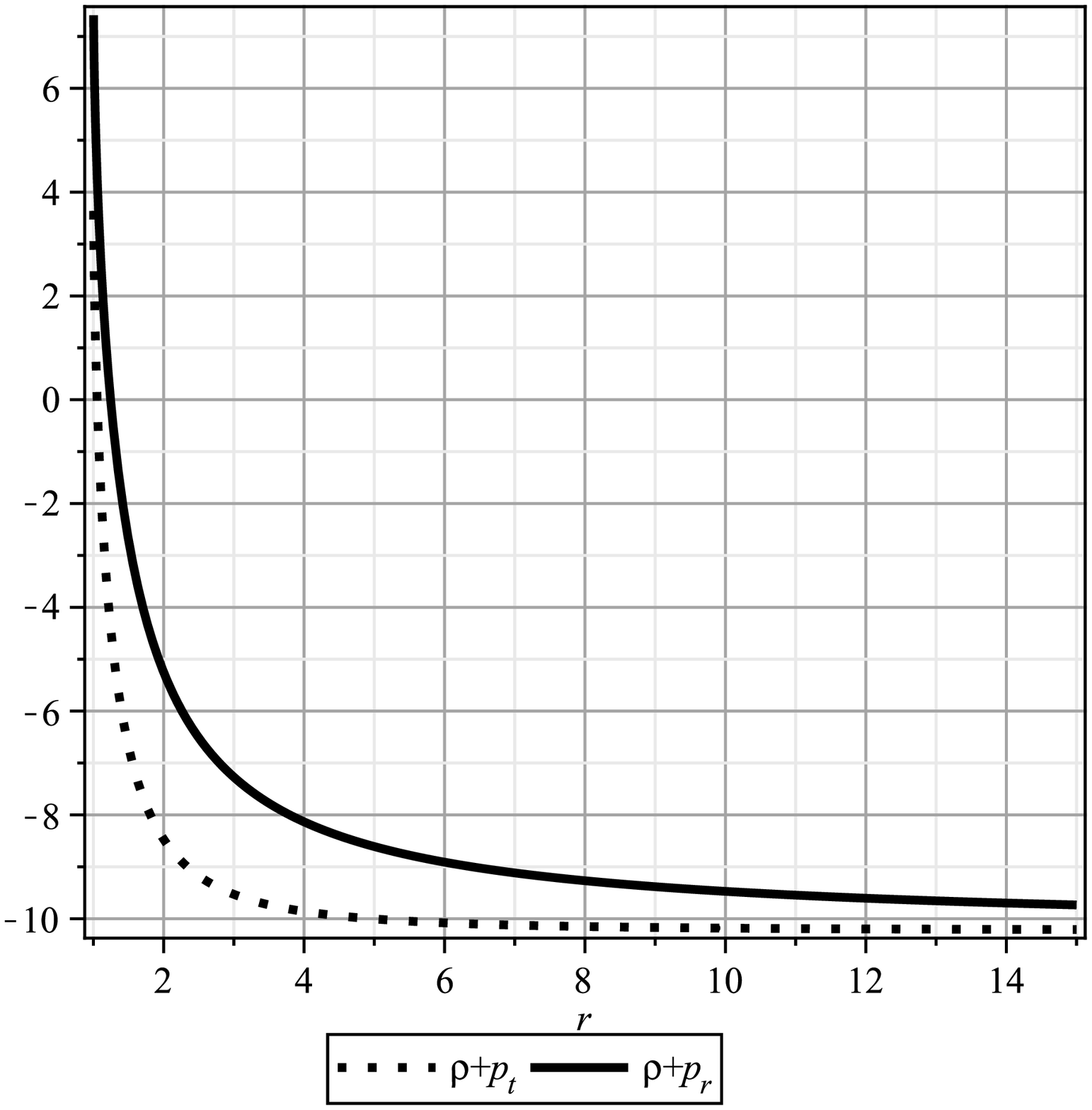} \\
\end{tabular}
\caption{ (\textit{Left}) The dotted curve shows the energy
density; the solid the radial pressure; and the
  dashed curve the tangential pressure. We have defined the following
  quantities: $r_0=1,n=0.2,c_1=6^{1/2}/2H_0(\Omega_{m0}-1),c_2=\Omega_{m0},\alpha=0$.
   (\textit{Right}) Variation of the $\rho+p_r$ solid , and $\rho+p_t$ dotted for the following
  quantities: $r_0=1,n=0.2,c_1=6^{1/2}/2H_0(\Omega_{m0}-1),c_2=\Omega_{m0},\alpha=0$.}
\end{figure*}

\section{Solution for Isotropic pressure}
To obtain the isotropic solutions we assume that
\begin{equation}
  a(r)=c, \qquad p_r(r)=p_t(r)=p(r).
\end{equation}
Now from
Eqs.(\ref{field:toff})-(\ref{field:thetaoff}), we obtain the
following solutions
\begin{eqnarray}
b _{\pm}( r ) &=&2\log  \Big[ \frac{-1\pm\sqrt {
-{r}^{2}C+1}}{1-{r}^{2}C\mp\sqrt { -{r}^{2}C+1}}\Big],
\end{eqnarray}
where $C$ is a constant of integration. The asymptotic behavior of
$\beta$ is $$\beta_{\pm}\simeq Cr^3+O(\frac{1}{r^2}),$$ which
diverges as $r\rightarrow\infty$. so the wormhole spacetime is not
asymptotically flat in both cases of $b_\pm$. Due to this reason, we
cannot define the ADM mass of wormhole.

To find out the radius of wormhole's throat, Following \cite{mattvisser}, we consider the
following line element
\begin{equation}
ds^2=e^{2\Phi(l)}dt^2-dl^2-r^2(l)d\Omega^2.
\end{equation}
To obtain the radius of throat, we define the proper distance
\begin{equation}
l(r)=\pm\int\limits_{r_0}^r\frac{dr'}{\sqrt{1-\frac{\beta_{\pm}(r')}{r'}}}.
\end{equation}
Solving the above integral, we obtain
$$l(r)=\mp\frac{\arcsin{(\sqrt{C}r})}{\sqrt{C}}.$$
Now by taking inverse on both sides
$$ r=\mp\frac{1}{\sqrt{C}}\sin{(l\sqrt{C})}. $$
Now the radius of throat $r_0=$min$\{r(l)\}$, is given by
$$|r_0|=\frac{1}{\sqrt{C}}.$$
Finally $C=\frac{1}{r_0^2}$.  By this solution we have the following
functions for the energy density $\rho$, and the pressure $p(r)$:
\begin{eqnarray}
4\pi \rho_{+} ( r ) &=&5/2\, \Big( -3/5\,{r_{{0}}}^{2}
 \Big( r \Big(  \Big( -16-16/3\,r \Big) {r_{{0}}}^{6}+ \Big( 28
\,{r}^{2}+{\frac {20}{3}}\,{r}^{3} \Big) {r_{{0}}}^{4}+ \Big( -5/3
\,{r}^{5}-13\,{r}^{4} \Big) {r_{{0}}}^{2}+{r}^{6} \Big) c_{{1}}
\sqrt {2}\nonumber\\&&+ \Big( -8/3\,c_{{2}}{r}^{2}-{\frac {128}{3}}\,\alpha
 \Big) {r_{{0}}}^{6}+10/3\,{r}^{2} \Big( {\frac {136}{5}}\,\alpha+c
_{{2}}{r}^{2} \Big) {r_{{0}}}^{4}-5/6\,{r}^{4} \Big( c_{{2}}{r}^{2}
+{\frac {344}{5}}\,\alpha \Big) {r_{{0}}}^{2}+{\frac {29}{3}}\,
\alpha\,{r}^{6} \Big)\nonumber\\&&\times \sqrt {-{\frac {{r}^{2}}{{r_{{0}}}^{2}}}+1}+
 \Big( -1/5\,rc_{{1}} \Big(  \Big( 16\,r+48 \Big) {r_{{0}}}^{4}+
 \Big( -12\,{r}^{3}-60\,{r}^{2} \Big) {r_{{0}}}^{2}\nonumber\\&&+{r}^{4} \Big(
15+r \Big)  \Big) {r_{{0}}}^{2}\sqrt {2}+ \Big( -8/5\,c_{{2}}{r}^
{2}-{\frac {128}{5}}\,\alpha \Big) {r_{{0}}}^{6}+6/5\,{r}^{2}
 \Big( c_{{2}}{r}^{2}+{\frac {104}{3}}\,\alpha \Big) {r_{{0}}}^{4}\nonumber\\&&-
1/10\,{r}^{4} \Big( c_{{2}}{r}^{2}+168\,\alpha \Big) {r_{{0}}}^{2}+
\alpha\,{r}^{6} \Big)  \Big( r+r_{{0}} \Big)  \Big( -r_{{0}}+r
 \Big)  \Big) {\frac {1}{\sqrt {-{\frac {{r}^{2}}{{r_{{0}}}^{2}}}+
1}}}{r_{{0}}}^{-6}{r}^{-2}\nonumber\\&&\times \Big( -2\,{r_{{0}}}^{2}-2\,\sqrt {-{\frac
{{r}^{2}}{{r_{{0}}}^{2}}}+1}{r_{{0}}}^{2}+{r}^{2} \Big) ^{-1}
 \Big( 1+\sqrt {-{\frac {{r}^{2}}{{r_{{0}}}^{2}}}+1} \Big) ^{-3},
\end{eqnarray}
\begin{eqnarray}
4\pi p_{+} ( r ) &=&-1/2\, \Big(  \Big( 20\,{r}^{2}{r
_{{0}}}^{4}-5\,{r}^{4}{r_{{0}}}^{2}-16\,{r_{{0}}}^{6} \Big) \sqrt {-
{\frac {{r}^{2}}{{r_{{0}}}^{2}}}+1}-13\,{r}^{4}{r_{{0}}}^{2}+{r}^{6}+
28\,{r}^{2}{r_{{0}}}^{4}-16\,{r_{{0}}}^{6} \Big)\nonumber\\&&\times  \Big( c_{{1}}
\sqrt {2}{r_{{0}}}^{2}+r \Big( 1/2\,c_{{2}}{r_{{0}}}^{2}+\alpha
 \Big)  \Big) {\frac {1}{\sqrt {-{\frac {{r}^{2}}{{r_{{0}}}^{2}}}+
1}}}{r_{{0}}}^{-6}{r}^{-1} \Big( -2\,{r_{{0}}}^{2}-2\,\sqrt {-{\frac
{{r}^{2}}{{r_{{0}}}^{2}}}+1}{r_{{0}}}^{2}+{r}^{2} \Big) ^{-1}\nonumber\\&&\times
 \Big( 1+\sqrt {-{\frac {{r}^{2}}{{r_{{0}}}^{2}}}+1} \Big) ^{-3}.
\end{eqnarray}
So the expression for the NEC reads by the following expression:
\begin{eqnarray}
\rho_{+}+p_{+}&=&1/2\, \Big( -3/4\, \Big( c_{{1}}r \Big(  \Big( -16/3\,r-{\frac {64
}{3}} \Big) {r_{{0}}}^{6}+{\frac {20}{3}}\,{r}^{2} \Big( r+{\frac {
26}{5}} \Big) {r_{{0}}}^{4}+ \Big( -5/3\,{r}^{5}-{\frac {44}{3}}\,{
r}^{4} \Big) {r_{{0}}}^{2}+{r}^{6} \Big) \sqrt {2}\nonumber\\&&+ \Big( -{
\frac {128}{3}}\,\alpha-16/3\,c_{{2}}{r}^{2} \Big) {r_{{0}}}^{6}+{
\frac {20}{3}}\,{r}^{2} \Big( c_{{2}}{r}^{2}+{\frac {64}{5}}\,\alpha
 \Big) {r_{{0}}}^{4}-5/3\,{r}^{4} \Big( c_{{2}}{r}^{2}+{\frac {152}
{5}}\,\alpha \Big) {r_{{0}}}^{2}+8\,\alpha\,{r}^{6} \Big) {r_{{0}}
}^{2}\nonumber\\&&\times\sqrt {-{\frac {{r}^{2}}{{r_{{0}}}^{2}}}+1}+ \Big( r+r_{{0}}
 \Big)  \Big( -r_{{0}}+r \Big)  \Big( -1/4\,c_{{1}}r \Big(
 \Big( 16\,r+64 \Big) {r_{{0}}}^{4}+ \Big( -72\,{r}^{2}-12\,{r}^{3
} \Big) {r_{{0}}}^{2}\nonumber\\&&+{r}^{4} \Big( 16+r \Big)  \Big) {r_{{0}}}
^{2}\sqrt {2}+ \Big( -32\,\alpha-4\,c_{{2}}{r}^{2} \Big) {r_{{0}}}^
{6}+ \Big( 48\,\alpha\,{r}^{2}+3\,c_{{2}}{r}^{4} \Big) {r_{{0}}}^{4
}+ \Big( -18\,{r}^{4}\alpha-1/4\,{r}^{6}c_{{2}} \Big) {r_{{0}}}^{2}
\nonumber\\&&+\alpha\,{r}^{6} \Big)  \Big) {\pi }^{-1}{\frac {1}{\sqrt {-{
\frac {{r}^{2}}{{r_{{0}}}^{2}}}+1}}}{r_{{0}}}^{-6}{r}^{-2} \Big( -2\,
{r_{{0}}}^{2}-2\,\sqrt {-{\frac {{r}^{2}}{{r_{{0}}}^{2}}}+1}{r_{{0}}}^
{2}+{r}^{2} \Big) ^{-1} \Big( 1+\sqrt {-{\frac {{r}^{2}}{{r_{{0}}}^
{2}}}+1} \Big) ^{-3}
\label{nec3}.
\end{eqnarray}
The qualitative behavior of the energy density and  the  pressure
for positive branch is plotted in Fig.2, left panel. Note that the
energy density remains positive. So the isotropic case satisfies the
WEC. Further, as we observe in the right panel, also the  NEC
satisfies here by the EM components.  The NEC is satisfied for both
cases of the radial and the tangential components of the pressure.
So there is at least one possibility to have a physically reasonable
wormhole solution with isotropic pressures in our viable $f(T)$
model. Finally in the isotropic case, for negative branch, from
figures 3, both null and weak energy conditions are violated.

For minus branch:
\begin{eqnarray}
4\,\pi \,\rho_{-} ( r ) &=&5/2\, \Big( 3/5\,{r_{{0}}}^{2}
 \Big( c_{{1}}r \Big(  \Big( -16/3\,r-16 \Big) {r_{{0}}}^{6}+
 \Big( {\frac {20}{3}}\,{r}^{3}+28\,{r}^{2} \Big) {r_{{0}}}^{4}+
 \Big( -13\,{r}^{4}-5/3\,{r}^{5} \Big) {r_{{0}}}^{2}+{r}^{6}
 \Big) \sqrt {2}\nonumber\\&&+ \Big( -8/3\,c_{{2}}{r}^{2}-{\frac {128}{3}}\,
\alpha \Big) {r_{{0}}}^{6}+10/3\,{r}^{2} \Big( c_{{2}}{r}^{2}+{
\frac {136}{5}}\,\alpha \Big) {r_{{0}}}^{4}-5/6\, \Big( c_{{2}}{r}^
{2}+{\frac {344}{5}}\,\alpha \Big) {r}^{4}{r_{{0}}}^{2}+{\frac {29}{
3}}\,\alpha\,{r}^{6} \Big)\nonumber\\&&\times \sqrt {-{\frac {{r}^{2}}{{r_{{0}}}^{2}}}+
1}+ \Big( r+r_{{0}} \Big)  \Big( -1/5\,c_{{1}}{r_{{0}}}^{2}r
 \Big(  \Big( 48+16\,r \Big) {r_{{0}}}^{4}+ \Big( -60\,{r}^{2}-12
\,{r}^{3} \Big) {r_{{0}}}^{2}\nonumber\\&&+{r}^{4} \Big( 15+r \Big)  \Big)
\sqrt {2}+ \Big( -{\frac {128}{5}}\,\alpha-8/5\,c_{{2}}{r}^{2}
 \Big) {r_{{0}}}^{6}+6/5\, \Big( c_{{2}}{r}^{2}+{\frac {104}{3}}\,
\alpha \Big) {r}^{2}{r_{{0}}}^{4}\nonumber\\&&-1/10\,{r}^{4} \Big(
168\,\alpha+c _{{2}}{r}^{2} \Big) {r_{{0}}}^{2}+\alpha\,{r}^{6}
\Big)  \Big( -r _{{0}}+r \Big)  \Big) {\frac {1}{\sqrt {-{\frac
{{r}^{2}}{{r_{{0}}
}^{2}}}+1}}}{r_{{0}}}^{-6}{r}^{-2}\nonumber\\&&\times \Big(
-2\,{r_{{0}}}^{2}+2\,\sqrt { -{\frac
{{r}^{2}}{{r_{{0}}}^{2}}}+1}{r_{{0}}}^{2}+{r}^{2} \Big) ^{-1 } \Big(
-1+\sqrt {-{\frac {{r}^{2}}{{r_{{0}}}^{2}}}+1} \Big) ^{-3},
\end{eqnarray}
\begin{eqnarray}
4\,\pi \,p_{-} ( r )&=&-1/2\, \Big( c_{{1}}\sqrt {2}{r_{{0
}}}^{2}+r \Big( 1/2\,c_{{2}}{r_{{0}}}^{2}+\alpha \Big)  \Big)
 \Big(  \Big( 16\,{r_{{0}}}^{6}-20\,{r}^{2}{r_{{0}}}^{4}+5\,{r}^{4}{
r_{{0}}}^{2} \Big) \sqrt {-{\frac {{r}^{2}}{{r_{{0}}}^{2}}}+1}-13\,{
r}^{4}{r_{{0}}}^{2}\nonumber\\&&+{r}^{6}+28\,{r}^{2}{r_{{0}}}^{4}-16\,{r_{{0}}}^{6}
 \Big) {\frac {1}{\sqrt {-{\frac {{r}^{2}}{{r_{{0}}}^{2}}}+1}}}{r_{{0
}}}^{-6}{r}^{-1} \Big( -2\,{r_{{0}}}^{2}+2\,\sqrt {-{\frac
{{r}^{2}}{ {r_{{0}}}^{2}}}+1}{r_{{0}}}^{2}+{r}^{2} \Big) ^{-1}
\nonumber\\&&\times\Big( -1+\sqrt {-{\frac
{{r}^{2}}{{r_{{0}}}^{2}}}+1} \Big) ^{-3},
\end{eqnarray}
and the NEC reads
\begin{eqnarray}
\rho_{-}+p_{-}&=&1/2\, \Big( 3/4\,{r_{{0}}}^{2} \Big( c_{{1}}r \Big(  \Big( -{
\frac {64}{3}}-16/3\,r \Big) {r_{{0}}}^{6}+{\frac {20}{3}}\, \Big(
r+{\frac {26}{5}} \Big) {r}^{2}{r_{{0}}}^{4}+ \Big( -{\frac {44}{3}
}\,{r}^{4}-5/3\,{r}^{5} \Big) {r_{{0}}}^{2}+{r}^{6} \Big) \sqrt {2
}\nonumber\\&&+ \Big( -{\frac {128}{3}}\,\alpha-16/3\,c_{{2}}{r}^{2} \Big) {r_{{0
}}}^{6}+{\frac {20}{3}}\,{r}^{2} \Big( {\frac {64}{5}}\,\alpha+c_{{2}
}{r}^{2} \Big) {r_{{0}}}^{4}-5/3\, \Big( c_{{2}}{r}^{2}+{\frac {152
}{5}}\,\alpha \Big) {r}^{4}{r_{{0}}}^{2}+8\,\alpha\,{r}^{6} \Big)\nonumber\\&&\times
\sqrt {-{\frac {{r}^{2}}{{r_{{0}}}^{2}}}+1}+ \Big( r+r_{{0}} \Big)
 \Big( -1/4\,{r_{{0}}}^{2} \Big(  \Big( 16\,r+64 \Big) {r_{{0}}}^
{4}+ \Big( -12\,{r}^{3}-72\,{r}^{2} \Big) {r_{{0}}}^{2}\nonumber\\&&+{r}^{4}
 \Big( 16+r \Big)  \Big) c_{{1}}r\sqrt {2}+ \Big( -4\,c_{{2}}{r}
^{2}-32\,\alpha \Big) {r_{{0}}}^{6}+ \Big( 3\,c_{{2}}{r}^{4}+48\,{r
}^{2}\alpha \Big) {r_{{0}}}^{4}\nonumber\\&&+ \Big( -1/4\,{r}^{6}c_{{2}}-18\,{r}
^{4}\alpha \Big) {r_{{0}}}^{2}+\alpha\,{r}^{6} \Big)  \Big( -r_{{0
}}+r \Big)  \Big) {\pi }^{-1}{\frac {1}{\sqrt {-{\frac {{r}^{2}}{{
r_{{0}}}^{2}}}+1}}}{r_{{0}}}^{-6}{r}^{-2}\nonumber\\&& \Big( -2\,{r_{{0}}}^{2}+2\,
\sqrt {-{\frac {{r}^{2}}{{r_{{0}}}^{2}}}+1}{r_{{0}}}^{2}+{r}^{2}
 \Big) ^{-1} \Big( -1+\sqrt {-{\frac {{r}^{2}}{{r_{{0}}}^{2}}}+1}
 \Big) ^{-3}.
\end{eqnarray}

\begin{figure*}[thbp]
\begin{tabular}{rl}
\includegraphics[width=7.5cm]{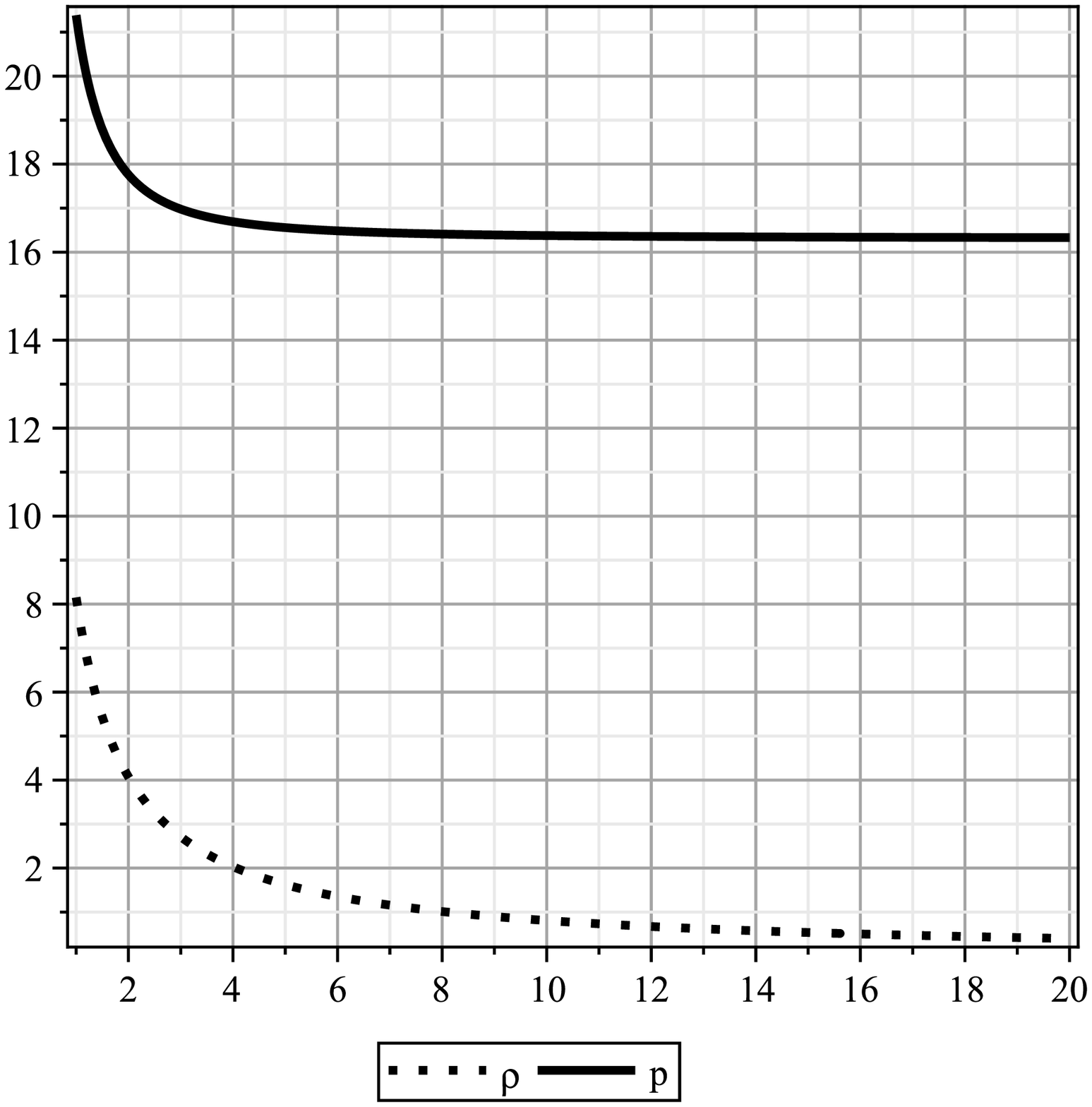}&
\includegraphics[width=7.5cm]{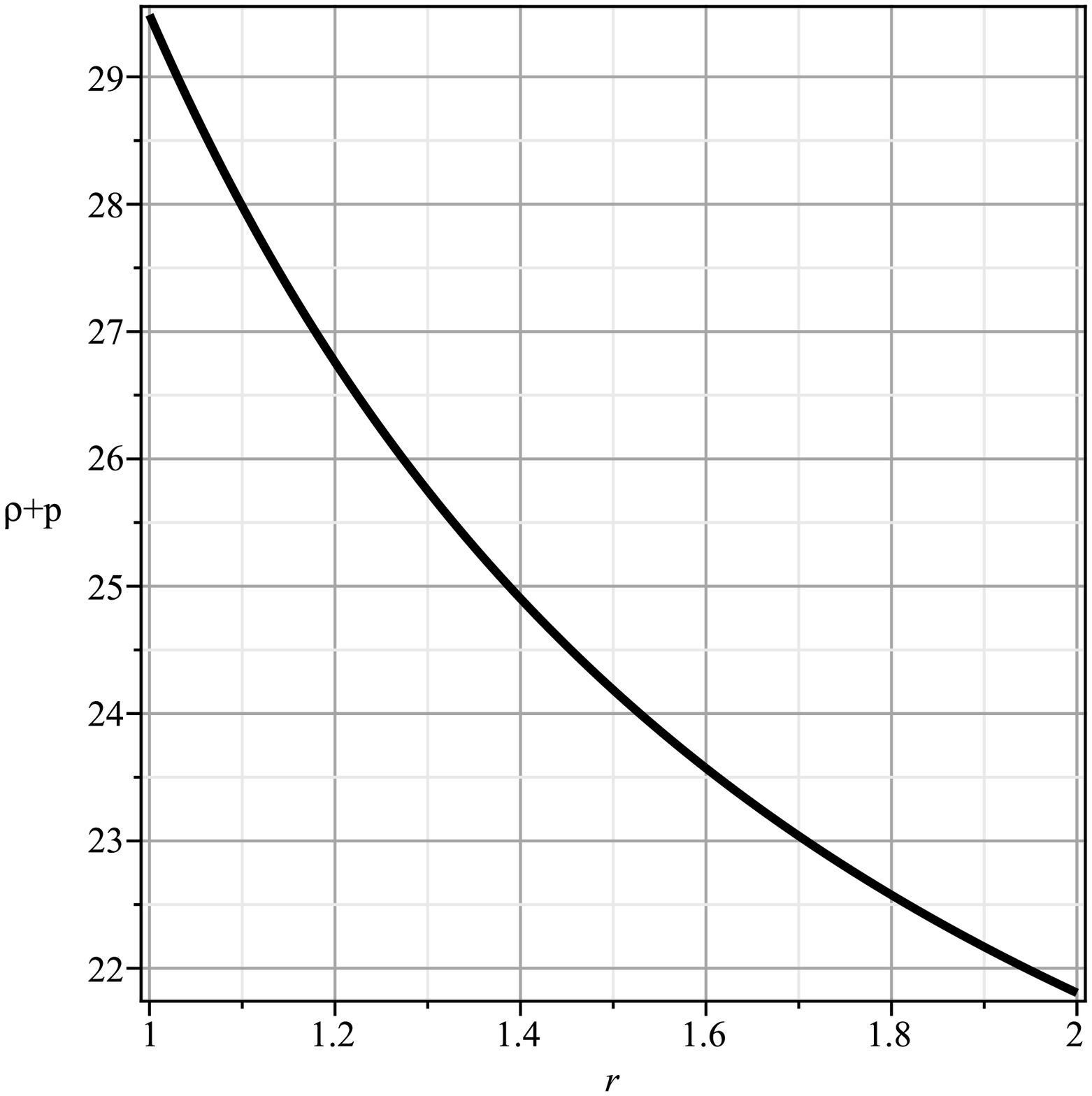} \\
\end{tabular}
\caption{ (\textit{Left})Positive Branch: The dotted curve for the energy
density; the solid as radial pressure. We have defined the
following
  quantities: $r_0=1,n=0.2,c_1=6^{\frac{1}{2}}/2H_0(\Omega_{m0}-1),c_2=\Omega_{m0},\alpha=0$.
  (\textit{Right}) Variation of the $\rho+p_r$ for the following
  quantities: $r_0=1,n=0.2,c_1=6^{\frac{1}{2}}/2H_0(\Omega_{m0}-1),c_2=\Omega_{m0},\alpha=0$.}
\end{figure*}

\begin{figure*}[thbp]
\begin{tabular}{rl}
\includegraphics[width=7.5cm]{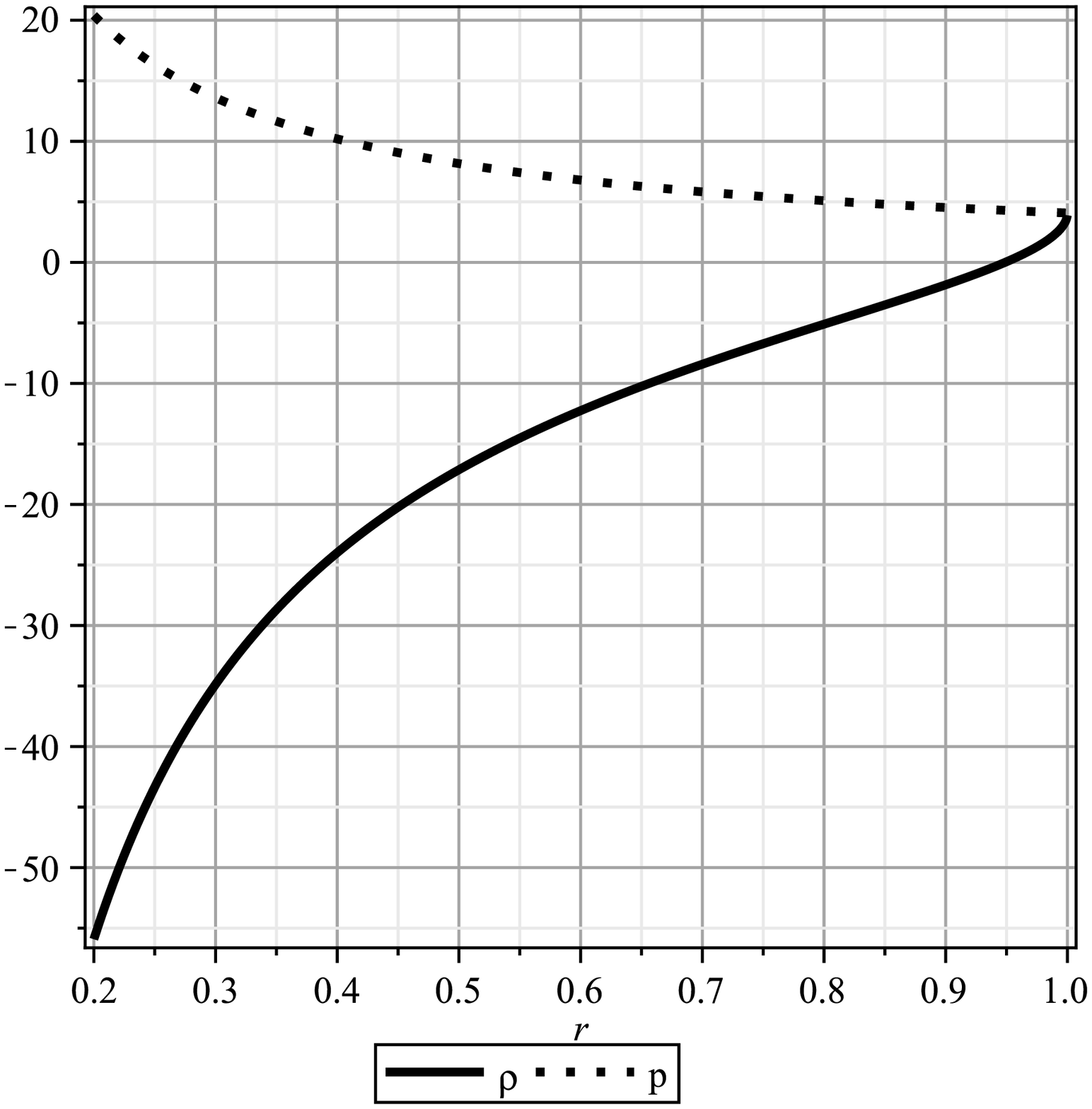}&
\includegraphics[width=7.5cm]{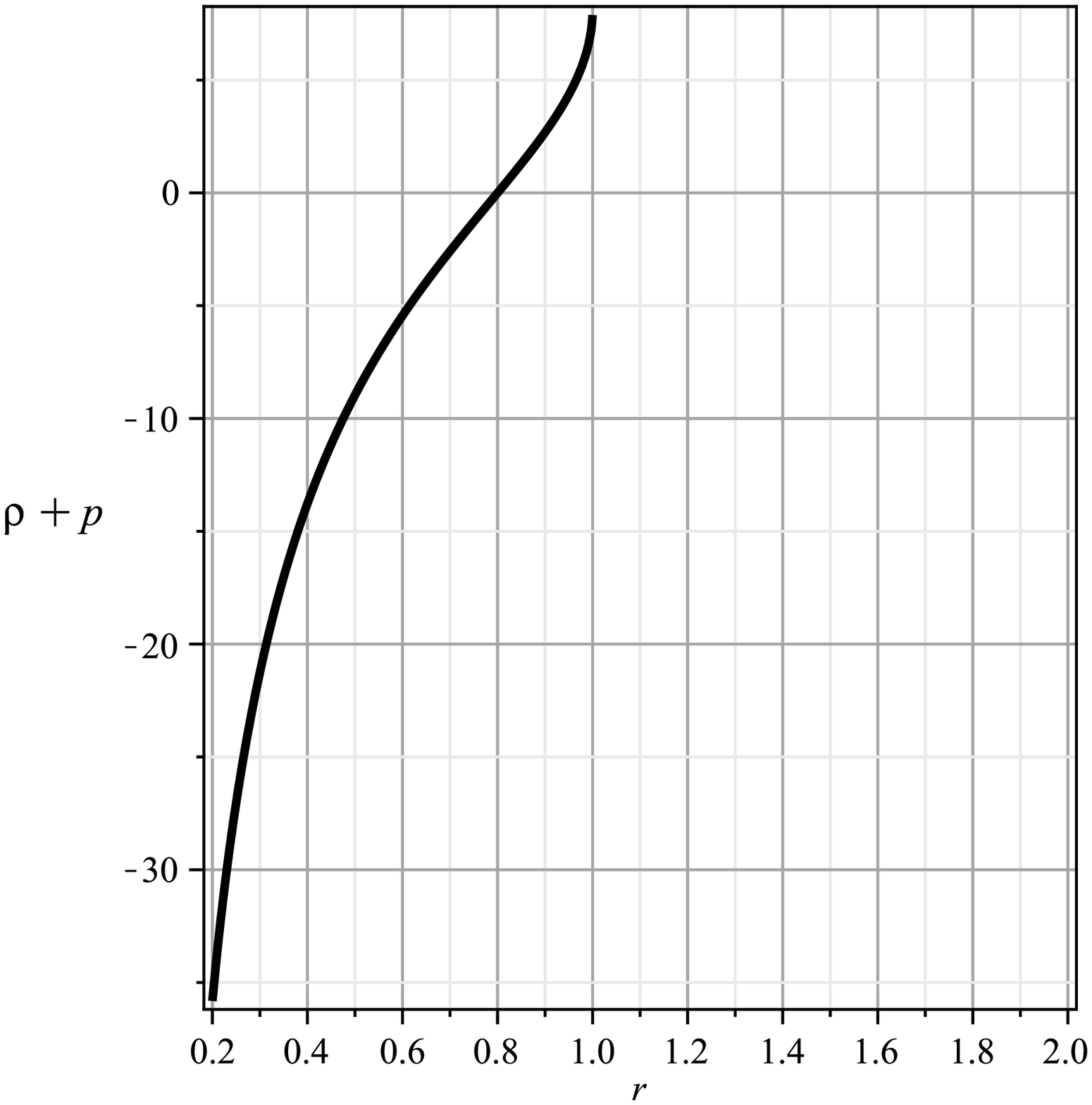} \\
\end{tabular}
\caption{ (\textit{Left}) Minus branch:The solid curve is for the energy
density; the dotted indicates the radial pressure. We have defined the
following
  quantities: $r_0=1,n=0.2,c_1=6^{\frac{1}{2}}/2H_0(\Omega_{m0}-1),c_2=\Omega_{m0},\alpha=0$.
  (\textit{Right}) Variation of the $\rho+p_r$ for the following
  quantities: $r_0=1,n=0.2,c_1=6^{\frac{1}{2}}/2H_0(\Omega_{m0}-1),c_2=\Omega_{m0},\alpha=0$.}
\end{figure*}

\section{Solution with barotropic equation of state }

In this case the exact solution exists only for $\alpha=0$. We assume
\begin{equation}\label{c}
a(r)=c,\quad p_r(r)=k\rho(r),
\end{equation}
where $k$ is an arbitrary but finite constant. Using (\ref{c}) in
the equations of motion, we obtain
\begin{eqnarray}
b( r ) &=&2\log  \Big( {\frac {{r}^{2}}{{r}^{2}+C(1-3r+
3{r}^{2}-{r}^{3})}} \Big),\\
 4\pi \rho( r)& =&-{\frac {c_{{2}}{r}^{6}+2c_{{1}
}\sqrt {2}{r}^{5}}{4{r}^{6}}},\\
 4\pi p_r( r)& =&-k{\frac {c_{{2}}{r}^{6}+2c_{{1}
}\sqrt {2}{r}^{5}}{4{r}^{6}}},\\
 4\pi p_{{t}}( r )& =&\frac {1}{4{r}^{6} ( -1+r ) }(-c_{{2}}{r}^{7}+c_{{2}
}{r}^{6}+3C\sqrt {2}c_{{1}}{r}^{6}-c_{{1}}\sqrt {2}{r}^{6}-9C \sqrt
{2}c_{{1}}{r}^{5}\nonumber\\&&-2c_{{1}}\sqrt {2}{r}^{5}+9C\sqrt
{2}c_{{1} }{r}^{4}-3C\sqrt {2}c_{{1}}{r}^{3}).
\end{eqnarray}
So, the expression for the transverse null energy condition
$\rho+p_t>0$ reads
\begin{eqnarray}
\rho+p_t&=&\frac {1}{16\pi {r}^{3}
 ( 1-r ) }(2c_{{2}}{r}^{4}+(-2c_{{2}}+3c_{{1}}\sqrt {2
}-3C\sqrt {2}c_{{1}}){r}^{3}+9C\sqrt
{2}c_{{1}}{r}^{2}\nonumber\\&&-9C \sqrt {2}c_{{1}}{r}+3C\sqrt
{2}c_{{1}})\label{nec4}.
\end{eqnarray}
Again, the qualitative behavior of the energy density and  the
pressure is plotted in Fig.4,  left panel, in $\log$ scale. Note
that the energy density is positive.
So the isotropic case satisfies the WEC. Further, as we observe in
the right panel, the  NEC is satisfied by the EM tensor components.
 The non violation of the NEC is for
the tangential component of the pressure. So there exists  at least
one possibility to have a physically acceptable wormhole solution
with barotropic radial fluid in our viable $f(T)$ model. In fact, in
the barotropic case, obviously the NEC is satisfied entirely  the
spacetime. In figure-5, we plot the ratio of pressure and energy
density and show that these remain positive for asymptotic values of
$r$.

\begin{figure*}[thbp]
\begin{tabular}{rl}
\includegraphics[width=7.5cm]{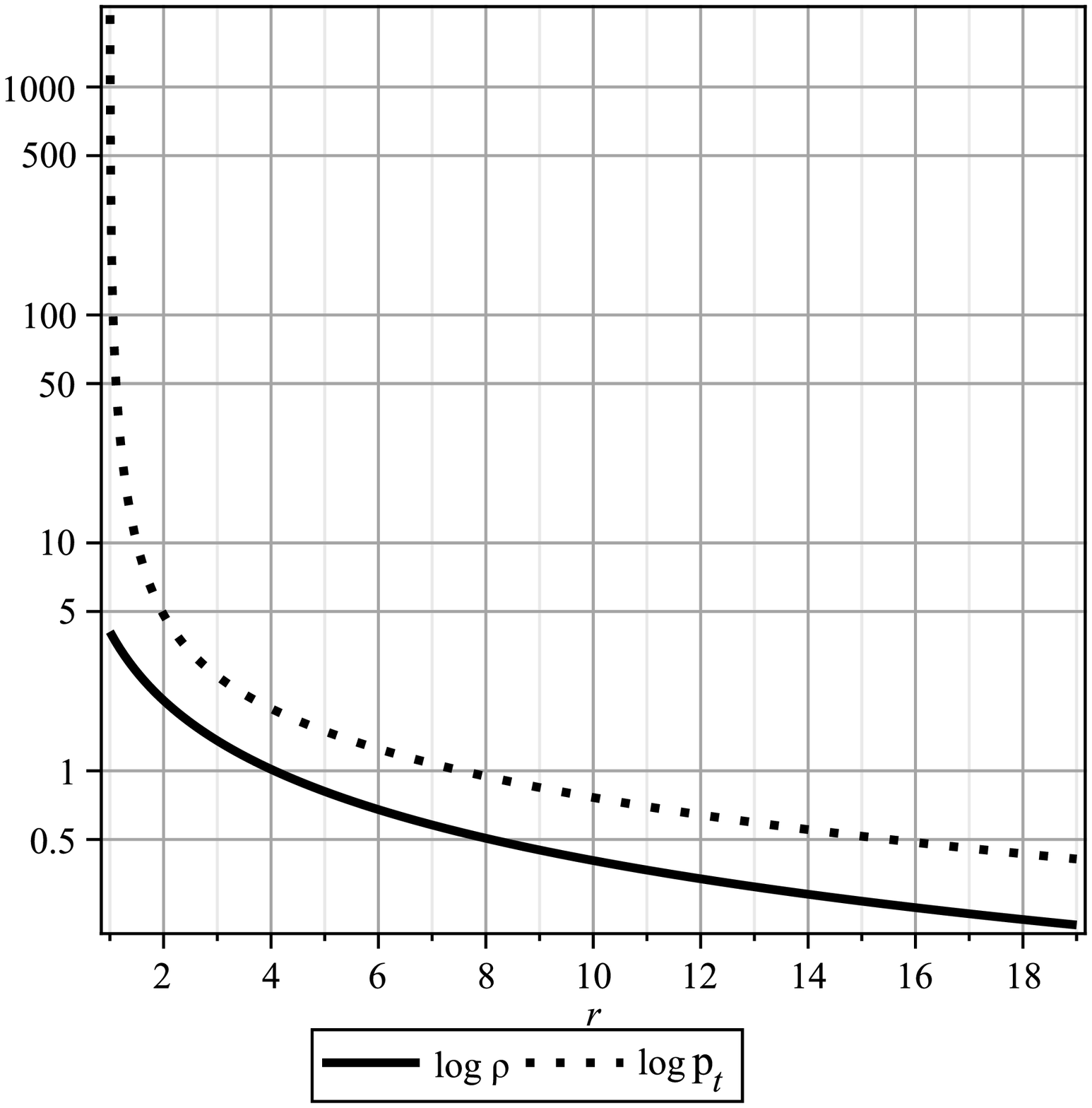}&
\includegraphics[width=7.5cm]{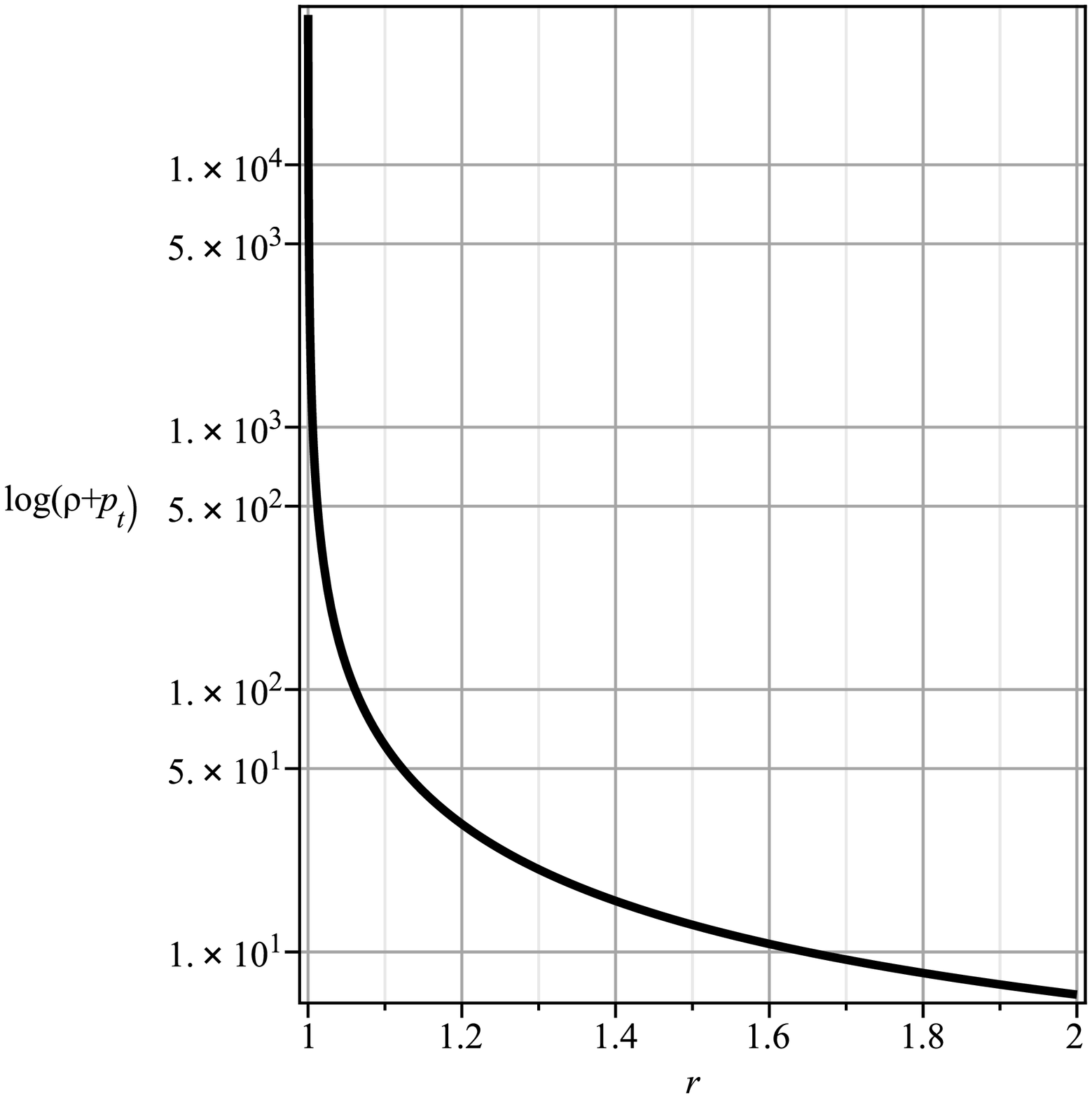} \\
\end{tabular}
\caption{ (\textit{Left}) We have defined the following
  quantities: $r_0=1,n=0.2,c_1=6^{\frac{1}{2}}/2H_0(\Omega_{m0}-1),c_2=\Omega_{m0},\alpha=0$.
  (\textit{Right}) Variation of the $\log(\rho+p_t)$ for the following
  quantities: $r_0=1,n=0.2,c_1=6^{\frac{1}{2}}/2H_0(\Omega_{m0}-1),c_2=\Omega_{m0},\alpha=0$.}
\end{figure*}

\begin{figure*}[thbp]
\begin{tabular}{rl}
\includegraphics[width=7.5cm]{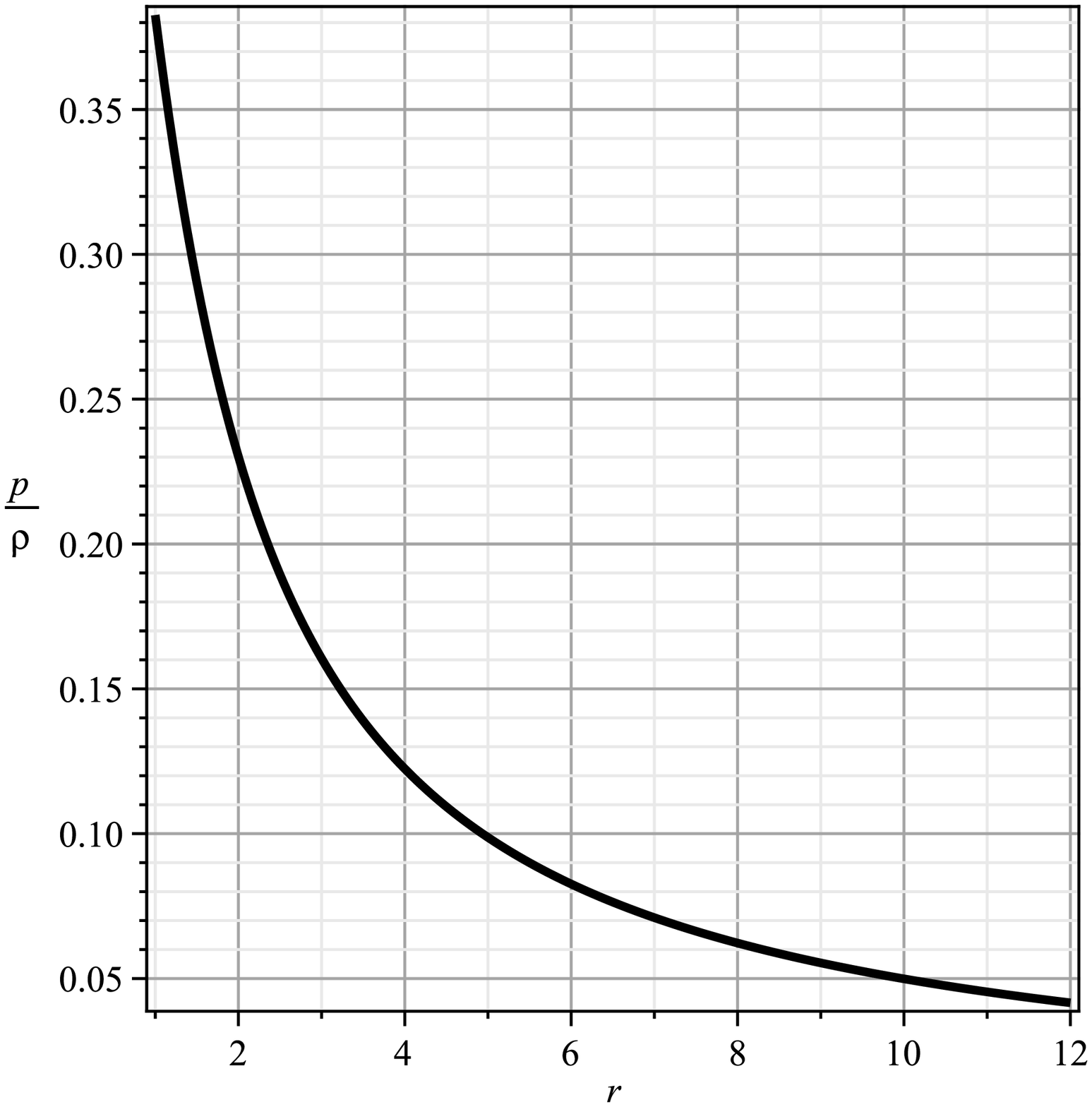}&
\includegraphics[width=7.5cm]{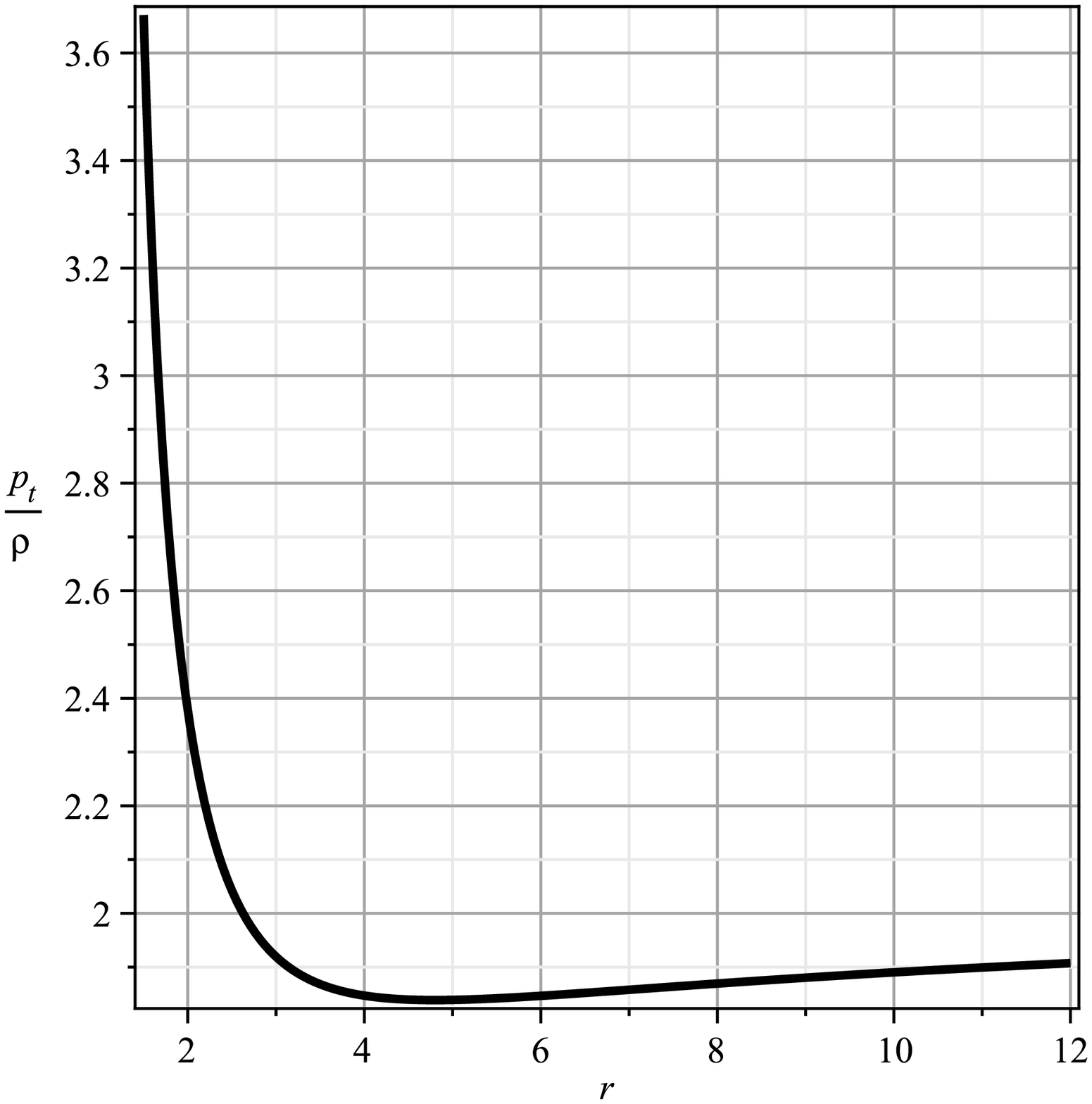} \\
\end{tabular}
\caption{ (\textit{Left}) The  curve depicts the EoS parameter $w=\frac{p}{\rho}$ for isotropic case. We have defined the following
  quantities: $r_0=1,n=0.2,c_1=6^{1/2}/2H_0(\Omega_{m0}-1),c_2=\Omega_{m0},\alpha=0$.
   (\textit{Right}) Variation of the EoS parameter $w=\frac{p_t}{\rho}$ for barotropic   case for the following
  quantities: $r_0=1,n=0.2,c_1=6^{1/2}/2H_0(\Omega_{m0}-1),c_2=\Omega_{m0},\alpha=0$.}
\end{figure*}

\section{Conclusion}

We have considered Morris -Thorne wormholes, i.e., static and
spherically symmetric traversable wormholes, in the Weitzenbock
spacetime with torsion. Basically, we discussed the possible
wormhole solutions in a viable $f(T)$ model with form
$f(T)=2c_1\sqrt{-T}+\alpha T+c_2$. This  model was picked up for its
simplicity in the numerical computation and astrophysical viability.
We investigated three kinds of the fluids including isotropic,
anisotropic and finally the barotropic fluids. We presented specific
solutions with various choices of the shape function. In all cases,
we obtained the exact solutions which described the wormhole
geometries. By checking the behavior of the weak and null energy
conditions for each case, we observed their violation for
anisotropic case while their satisfaction for isotropic and
barotropic cases. So we can have both isotropic and barotropic
wormhole solutions in this viable torsion based model of the
gravity.

In anisotropic case, our $f(T)$ model mimics the phantom energy
since both energy conditions NEC and WEC are violated $r>r_0$. In
isotropic case, we have two special cases for shape function. For
positive branch, both energy conditions are satisfied while they are
violated in the negative branch case. Also the obtained wormhole
solution is not asymptotically flat. For barotropic case, again we
have non-asymptotically flat solution and energy conditions are
satisfied for transverse NEC and WEC. Moreover, we discussed the
behavior of the EoS parameter $w=p/\rho$ for isotropic and
barotropic cases. Our numerical simulation shows that for isotropic
case, $w$ remains positive. While for barotropic case, the same
behavior happens.

\end{document}